\documentclass[PRL,twocolumn]{revtex4}
\usepackage{amsmath,amssymb,latexsym,epsfig,graphics,epsf}
\usepackage{fixltx2e} 
\usepackage{graphics}
\usepackage{float}
\usepackage{graphicx}
\usepackage{caption}
\usepackage{multirow}
\usepackage{subfigure}

\newcommand{\scA}{\mbox{$\mathcal{A}$}}

\newcommand{\be}{\begin{equation}}
\newcommand{\ee}{\end{equation}}
\begin{document}
\title{`It from Bit': is there a second law of quantum complexity?}
\author{S.\ Davatolhagh}
\altaffiliation{Corresponding Author: davatolhagh@shirazu.ac.ir}
\author{A.\ Sheykhi}
\author{M.H.\ Zarei}
\affiliation{Department of Physics, School of Science, Shiraz
University, Shiraz 71946, Iran}
\date{\today}
\begin{abstract}

At a deeper level the principle of least action is interpreted as
the law of {\em least entropy increase} consistent with
Prigogine's principle of minimum entropy production, and the
implications of quasistatic information quantization rule (Proc.\
R.\ Soc.\ A {\bf 480}: 20240024), are explored for the conjectured
second law of quantum complexity. It is thus shown that the
conjectured second law of complexity is derivable from the
information quantization rule such that long after heat-death the
quantum state complexity evolves as $C(t)=C_{\rm max}\exp(-1/t)$,
increasing with time to a saturation value $C_{\rm max}$ that is
exponential in the equilibrium entropy $S_{\rm max}$. For the
out-of-equilibrium circumstances, however, the quantum complexity
can decrease with the time, asymptotically tending to a minimum
determined by the distance from equilibrium.

\end{abstract}
\pacs{}
\maketitle


The classical law of the increase of entropy, also known as the
second law of thermodynamics, appears to miss a significant part
of the directed evolution as systems continue to evolve at a
quantum level long after thermal equilibrium. As the number of
orthogonal quantum states in Hilbert space increases exponentially
with the number of qubits, the system starting from a simple
quantum state will take an enormously long time to explore the
entire space of states (including the superpositions), much longer
than what is needed to establish thermodynamic equilibrium. This
phenomenon, sometimes referred to as `life after heat-death', was
first encountered as a black hole paradox such that the interior
volume and the gravitational action of the causally connected
Wheeler-DeWitt patch of a black hole interior seemed to grow
indefinitely long after the black hole had reached the state of
maximum entropy, thermal equilibrium, or heat-death. In solving
this paradox Susskind and colleagues reached an explanation in
terms of holography linking the volume as well as the
Wheeler-DeWitt action of the black hole's interior with the
complexity of the boundary quantum state \cite{CV,CA}. So the
second law of quantum complexity was proposed in direct analogy to
the second law of classical thermodynamics \cite{BS}. Although it
is not a law in the sense that it has been proven, but it is
conjectured that such a law must indeed exist although the extent
to which it applies to systems beyond black holes is not entirely
clear.

In the context of statistical thermodynamics, the entropy of a
system is given by \be S  = -k_{\rm B} \sum_i p_i \ln p_i =
-k_{\rm B} \langle \ln p_i \rangle \ee where, $p_i$ is the
probability of a microstate $i$, and $k_{\rm B}$ is the Boltzmann
constant. Entropy $S$ is therefore the ensemble average of a
quantity called the surprisal $s_i=\ln (1/p_i)$, which is a
measure of how surprising or informative is the realization of a
microstate (or an event) with probability $p_i$ \cite{Shannon}.
The more the ensemble average of surprisal, the more is the
entropy or missing information. From this Shannon interpretation
of entropy as the missing information, a natural definition for
information available arises as the lack of entropy (or the lack
of missing information) that was first proposed by David Layzer
\cite{Layzer}, and later elaborated on by Eric Chaisson and others
\cite{Chaisson,DSZ}: \be \Delta S(t) \equiv S_{\rm max}(t) - S(t).
\ee In Eq.\ (2), $S_{\rm max}(t)$ is the maximum attainable
entropy (or the equilibrium entropy) of the thermodynamic system
consistent with the external constraints at time $t$, and $S(t)$
is the actual entropy of the system at the same time $t$. Clearly,
$\Delta S $ measures the lack of missing information and is
therefore the information available. It also embodies Layzer's
definition of order as the lack of disorder. There is an implicit
assumption of local equilibrium such that locally the entropy
$S(\vec{x},t)$ is well-defined, and $S(t)=\int S(\vec{x},t)\,d^3x
$ can be obtained by integrating over the thermodynamic system.
Furthermore, a {\em thermodynamic potential for information} can
be defined by $T(t)\Delta S(t)$ \cite{DSZ}, where $T(t)=
V^{-1}\int T(\vec{x},t) d^3x $ is the mean temperature of the
thermodynamic system. $T(t)\Delta S(t)$ is indeed a thermodynamic
potential (or free energy) as its minimization gives the
equilibrium state $S=S_{\rm max}$, and of course has the
dimensions of energy.

In the present work we begin by showing that the principle of
least action, can be expressed as a law of least entropy increase
-- similar to Prigogine's principle of minimum entropy production
for systems near thermodynamic equilibrium \cite{Prigogine} --
such that in view of Eq.\ (2) the physical systems have a natural
tendency to spend their information resource economically, and
then proceed to address the question of `is there a second law of
quantum complexity?', which has long puzzled the physicists
working on the evolution of black hole's interior volume and
gravitational action beyond thermal equilibrium \cite{CV,CA,BS},
as well as the computer scientists working on the problem of
measuring the computational complexity of quantum states or
computer circuits needed to prepare them
\cite{computer1,computer2}. This conjectured second law of quantum
complexity asserts that the quantum state of a chaotic system
continues to evolve and increase in complexity long after the
system has reached thermodynamic equilibrium until the
computational complexity of the quantum state reaches a maximum
value of order $\exp{(S_{\rm max})}$ determined by the equilibrium
entropy of the system $S_{\rm max}$. We explore the conditions
under which such a law must hold, and essentially derive it from
quasistatic information quantization rule developed in Ref.\
\cite{DSZ}, paying particular attention to the time evolution of
quantum complexity in the intermediate as well as long-time
regime. This analysis leads to falsifiable predictions for the
dynamics of quantum complexity that may be tested against model
calculations and simulations.

{\it Entropic interpretation of the laws of motion.}--- Although
it is widely believed that the laws of physics must have their
origin in information theory -- a conjecture best expressed by the
late Wheeler's catchphrase `It from Bit' \cite{Wheeler} -- there
has so far been limited progress in identifying the informational
basis of the laws of nature (see, however, \cite{Layzer, Chaisson,
DSZ}). If information is truly a fundamental quantity as Wheeler's
conjecture clearly suggests, then the fundamental laws of physics,
in particular the principle of least action, must indeed be
expressible as laws governing entropy. To show this, we employ the
original version of the principle of least action due to
Maupertuis, which is also derivable from Hamilton's principle
\cite{mechanics}.

Maupertuis' principle of least action $\delta\scA =0$, states that
with energy held fixed the classical path makes the action
stationary among all paths that connect the same initial and final
configuration points. Maupertuis' action is the line integral in
phase-space  between two configuration points $q_1$ and $q_2$
subject to constant energy constraint: \be \scA = \int p\,dq =
\int\sqrt{2m[E-V(q)]}\,dq,\ee where $q$ is the coordinate and
$p\equiv\sqrt{2m[E-V(q)]}$ the conjugate momentum. The action
$\scA$ of course depends on the path in configuration space taken
between initial and final configuration point. Allowing some
uncertainty in energy $\Delta E \ll E$ (as one always should), and
with fixed energy constraint ($E$\,,\,$E+\Delta E$), the
accessible phase-space points possibly swept out by the particle
are confined to a strip, which we call the {\em action strip},
given by \be \Delta\scA \equiv \scA(E+\Delta E) - \scA(E)\simeq
\sqrt{\frac{m}{2}}\Delta E\int\frac{dq}{\sqrt{E-V(q)}}\,.\ee

\begin{figure}[H]
\centering
\includegraphics[width =0.4\textwidth]{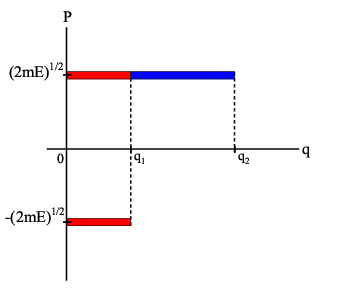}
\caption{The phase-space diagram of a free particle with energy
$E$ and uncertainty $\Delta E\ll E$ moving from configuration
point $q_1$ to $q_2$. The blue action strip represents the
shortest or the least action path between the two points while the
longer (blue + red) is another path that also touches the origin
in going from $q_1$ to $q_2$. Because the area of each action
strip in units of Planck's constant $h$ determines the number of
quantum states $\Omega$ possibly swept out by the particle in
going from $q_1$ to $q_2$, the least action strip (blue) is also
the least entropic as $S=K \log\Omega$. } \label{fig1}
\end{figure}

For the sake of illustration, we consider a free particle
($V(q)=0$) whose action strip by Eq.\ (4) becomes $\Delta\scA =
\sqrt{m/2E}\,\Delta E\, L$, where $L$ is the length of the path
traveled between initial and final configuration points $q_1$ and
$q_2$ as shown in Fig.\ 1 for the least action path (blue strip),
and another longer path that also touches the origin (red strips +
blue strip). Now because the area associated with every microstate
(or quantum state) in the phase-space is $\delta p \delta q \sim
h$, where $h$ is the Planck's constant, the number of microstates
$\Omega$ accessible to the particle in going from $q_1$ to $q_2$
is \be \Omega = \frac{\Delta\scA}{h}=
\frac{1}{h}\sqrt{\frac{m}{2E}}\,\Delta E\,L.\ee One can assign an
entropy to the path by the logarithmic number of accessible
quantum states or the logarithmic number of phase-space
microstates the particle could  cover in going from initial to
final configuration point. This is called the {\em action entropy}
\be S = K\log\Omega = K \log\left(\frac{1}{h}\sqrt{\frac{m}{2E}}\,
\Delta E\, L\right), \ee where constant $K$ can be adjusted to set
the unit or any desired base for the logarithm. Clearly the path
of least action, i.e.\ the path with the shortest length $L$, is
also the path of least entropy or uncertainty with regards to the
set of microstates covered in going from initial to final
configuration point. Because the extremum of action integral
$\scA=\sqrt{2mE}\,L$ also makes $\Delta\scA$ an extremum,
therefore, the Maupertuis' principle of least action $\delta\scA =
0$ picks out the path for which the phase-space strip is an
extremum, as a result of which the least action principle can be
reinterpreted as the principle of {\em least entropy increase} at
fixed energy ($E$\,,\,$E+\Delta E$).

From the point of view of path-integral formulation of quantum
mechanics \cite{FH}, all paths contribute to the transition
amplitude such that the amplitude  for going from a configuration
point $q_1$ at time $t_1$ to another point $q_2$ at time $t_2$ is
given by \be \langle q_2,t_2|q_1,t_1\rangle = \int D[q(t)]\,\exp
\left(i\scA[q(t)]/\hbar\right). \ee Every possible path $q(t)$
between the two points contributes a complex phase determined by
its action $\scA[q(t)]$, and the classical least action path
$\delta\scA = 0$ emerges from interference in the semiclassical
limit of small $h$ because paths with large actions tend to cancel
out due to wildly fluctuating phases. So from a slightly different
point of view where {\em action is viewed as a measure of path
distinguishability}, a bundle of paths with large action
differences (non-vanishing first variation) are highly entropic
thus interfere incoherently and cancel out. On the other hand the
bundle of paths around a classical path of stationary action
($\delta\scA =0$) are indistinguishable by action to first order,
and therefore least entropic resulting in coherent interference in
the semiclassical limit. So from the point of view of amplitude
mechanics as well, the classical principle of least action is in
fact at a more fundamental level a principle of least entropy
increase, akin to Prigogine's principle of least entropy
production for systems near thermodynamic equilibrium. In view of
Eq.\ (2), this speaks of the natural tendency of a physical system
(in this case a classical particle) to retain as much information
as possible by adopting a path that only increases the entropy
(missing information) by a bare minimum.

The action entropy discussed for a system moving in one dimension
can be readily generalized to higher $d$-dimensional cases where
the system is separable such that the potential energy can be
written in the form $V(\vec{q})=\sum_i V_i(q_i)$. Again
Maupertuis' action is a line integral in phase-space \be \scA(E) =
\int \vec{p}\cdot d\vec{q} =\sum_{i=1}^d\int
p_i\,dq_i=\sum_{i=1}^d \scA_i(E_i), \ee where the momentum
component $p_i \equiv\sqrt{2m[E_i - V_i(q_i)]}$. According to Eq.\
(8) the action is separable, much like the energy $E=\sum_iE_i$,
and the one-dimensional analysis presented before can be
separately applied to every component of motion. In particular,
the $d$ number of action strips in the $2d$-dimensional
phase-space all contribute to the action uncertainty $\Delta\scA$
such that \be \Delta\scA = \scA(E+\Delta E) - \scA(E) =
\sum_{i=1}^d \Delta\scA_i \ee where, similar to the
one-dimensional case \be \Delta\scA_i = \scA(E_i+\Delta E_i) -
\scA(E_i)\simeq
\frac{1}{2}\int\sqrt{\frac{2m}{E_i-V_i(q_i)}}\,\Delta
E_i\,dq_i.\ee $\Delta\scA$ in units of $h$ is the number of
quantum states accessible to the particle in moving from initial
to final configuration point given the fixed energy constraint
($E$ , $E+\Delta E$): \be \Omega = \frac{\Delta\scA}{h}=
\sum_{i=1}^d \Omega_i \ee where, $\Omega_i = \Delta\scA_i/h$, and
the action entropy becomes \be S = K\log\Omega =
K\log\left(\sum_{i=1}^d \Omega_i\right).\ee

It must be pointed out that although Hamilton's principle of least
action is formulated in terms of the Lagrangian function and uses
a slightly different action integral \be \scA = \int_{t_1}^{t_2}
L(q,\dot{q}) dt = \int_{t_1}^{t_2} (p\dot{q}-H)dt\,,\ee but with
the energy held fixed ($H=E$) it formally reduces to Maupertuis'
principle of least action such that the application of Hamilton's
principle to Eq.\ (13) and eliminating time as a variable leads to
a variational principle in the phase-space \cite{mechanics}. This
implies that the laws of physics that are formulated in terms of a
Lagrangian, in principle can be given a similar entropic or
information-theoretic interpretation as given here for the
principle of least action encountered in classical mechanics.

{\it Complexity and action.}--- The computational complexity
$C(t)$ of a quantum state is roughly the minimum number of quantum
gates required to prepare the state from a simple reference one.
The conjectured second law of complexity put forward by Brown and
Susskind proposes that the quantum state continues to become more
complex or harder-to-prepare computationally as the system
explores more of its Hilbert space long after it has reached
thermodynamic equilibrium, until it finally saturates to a maximum
given by the equilibrium entropy $C_{\rm max}\sim\exp(S_{\rm
max})$ \cite{BS}.

For a black hole and its dual boundary, the `Complexity = Volume'
(CV) conjecture proposes \be C(t)\sim\frac{V_{\rm int}}{G\,l_{\rm
AdS}}, \ee where $V_{\rm int}$ is the interior volume of the black
hole (as seen in a Penrose diagram), $G$ is Newton's constant, and
$l_{\rm AdS}$ is the AdS radius of the space-time \cite{CV}.
However, a more interesting proposal from the point of view of the
present work is `Complexity = Action' (CA), which instead
associates the boundary complexity at a given time $t$ with the
gravitational action of causally connected Wheeler-DeWitt (WDW)
space-time patch in the bulk \cite{CA}: \be
C(t)\sim\frac{\scA_{\rm WDW}}{\pi\hbar}.\ee As boundary time
increases, the WDW patch stretches deeper into the black hole
interior, thus increasing the complexity. CA is often considered
intuitively more attractive as it relates the dynamical history of
the gravitational system (encoded in WDW action) to its quantum
state complexity. The key advantage of the CA correspondence is
its {\em generalizability}  such that similar to a black hole, the
dynamical history of any system encoded in its accessible action
strip must determine its quantum state complexity. In other words,
the causally accessible action $\scA_{\rm WDW}$ of the
gravitational system and the accessible action strip $\Delta\scA$
of a mechanical system in units of Planck's constant $h$ both
specify the part of the state-space $\Omega$ swept out by the
respective system, and must therefore determine the quantum
complexity.

{\it Quasistatic information quantization rule.}--- This rather
recent quantization rule was developed in Ref.\ \cite{DSZ}, where
it was applied to different stages of cosmic evolution. It is
founded on (i) Wentzel-Kramers-Brillouin (WKB) semiclassical
energy quantization rule \cite{Sakurai}, and (ii) quasistatic
information-energy equivalence principle of Landauer
\cite{Landauer}, and Szilard before him \cite{Szilard}, both of
which have been proven empirically in the more recent years
\cite{exp2012,exp2010}.

In view of the semiclassical WKB approximation of the Schrodinger
eigenvalue equation, the action integral is quantized such that
\be \int_{x_1}^{x_2} p(x) dx = (n+\delta) \pi \hbar\ , \ee where
$x_1$ and $x_2$ are the classical turning points. By applying a
change of variables from position to time $\int_{x_1}^{x_2} p(x)
dx = \int_{t_1}^{t_2} p(x(t)) \dot{x} dt$, and using $p\equiv
\sqrt{2m(E-V(x(t))}$ where $E-V(x(t))=m\dot{x}^2/2$ is the
instantaneous kinetic energy, the energy quantization rule is
obtained as a time integral over the dynamical history: \be
\int_{t_1}^{t_2}\left[E_n - V(x(t))\right] dt =
\frac{1}{2}(n+\delta)\pi\hbar.\ \ee In Eq.\ (17), $E_n$ is the
semiclassical energy eigenvalue, $V(x(t))$ is the potential energy
well, $n=1,2,\cdots$ is a positive integer, $\delta$ is a
system-dependent fractional number in the range $-1<\delta < 1$,
and $\hbar = h/2\pi$. $x(t)$ is the classical least action
trajectory of the mechanical system, and the time limits $t_1$ and
$t_2$ correspond to the classical turning points when $E_n =
V(x(t))$.

Furthermore by invoking the {\em quasistatic information-energy
equivalence principle} \cite{Vopson, DSZ}, that is embodied in
both Landauer's quasistatic limit, which specifies the minimum
amount of heat generated by quasistatically erasing a bit of
information $Q({\rm 1\,bit}) = k_{\rm B}T\ln 2$ ($T$ being the
ambient temperature) \cite{Landauer, exp2012}, and a well-studied
theoretical model of a quasistatic information heat engine due to
Szilard that implies a bit of information can be quasistatically
converted to the mechanical work equivalent of the Landauer's
limit \cite{Szilard, exp2010}, the quasistatic information
quantization rule is thus obtained from the energy quantization
rule  Eq.\ (17) \cite{DSZ}:\be \int_{t_n}^{t}\left[I_n -
V(t)\right] dt = \tilde{n} \ee where, \be V(t)= -T(t)\Delta
S(t)\ee is the time-dependent thermodynamic potential well for
information, $t$ is the present time, $t_n$ is the time when the
information state first appears given by $I_n = V(t_n)$, and
$\tilde{n}=\frac{1}{2}(n+\delta)\hbar$. The quantized information
eigenvalue $|I_n|$ is interpreted as the information content of a
quasistatic information state supported by the out-of-equilibrium
environment (for further details the interested reader is invited
to read Ref.\ \cite{DSZ}).

{\it Derivation of the second law of quantum complexity.}--- We
begin by observing that the quantized eigen-information $|I_n|$ is
a measure of the information resource available to the quantum
state $n$ and must indeed relate to its uncomplexity $U_n$, which
is itself the resource available for quantum computation
\cite{BS}: \be U_n = C_{\rm max} -C_n.\ee  To obtain the exact
relationship between $|I_n|$ and $U_n$, consider that the
thermodynamic potential for information $T(S_{\rm max}- S)$
essentially measures the distance of the environment from
equilibrium. The quantized eigen-information $|I_n|$ derived from
it can similarly be written in the form of a distance $|I_n| = T
(S_{\rm max} - S_n)$ such that \be \frac{|I_n|}{T} = S_{\rm max} -
S_n ,\ee  where $S_n$ is the {\em complexity entropy} of the
information state \cite{complexityS}. This identification of $S_n$
with complexity entropy arises from the fact that the
eigen-information $|I_n|$ must indeed be a measure of uncomplexity
because the state of minimum eigen-information $|I_n| = 0$ implies
$S_n=S_{\rm max}$, which on exponentiation gives the state of
maximum complexity as it should: \be C_{\rm max} =
\exp\left(\frac{S_{\rm max}}{k_{\rm B}}\right).\ee The complexity
entropy $S_n$ is of course the logarithmic measure of complexity
$C_n$, so that by Eq.\ (21) \be C_n = \exp\left(\frac{S_n}{k_{\rm
B}}\right) = C_{\rm max}\exp\left(-\frac{|I_n|}{k_{\rm
B}T}\right),\ee which on substitution in Eq.\ (20) gives the
uncomplexity \be U_n = C_{\rm max} \left[ 1-
\exp\left(-\frac{|I_n|}{k_{\rm B}T}\right)\right].\ee

\begin{figure}[H]
\centering
\includegraphics[width =0.4\textwidth]{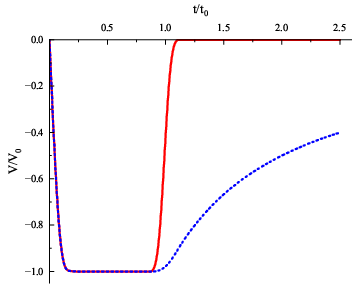}
\caption{The information potential well of Eq.\ (25) with $V(t)$
measured in units of the well depth $V_0$, and time in units of
the well width $t_0$, representing an environment
out-of-equilibrium for the time period $0<t<1$, returning to
equilibrium at $t=1$ with a time constant $\tau_{\rm eq}\ll 1$,
and staying in equilibrium for all times $t>1$. The blue dotted
curve shows $I_1(t)$. Clearly the highest eigen-information
$|I_1(t)|$ continues to evolve as $t^{-1}$ long after heat-death,
as all other states do (not shown). This is the part of evolution
that the second law of thermodynamics tends to miss, but is
captured by the second law of quantum complexity.} \label{fig2}
\end{figure}

To study the implications of quasistatic information quantization
rule Eq.\ (18) for the conjectured second law of quantum
complexity, consider the thermodynamic potential well for
information given in Eq.\ (25) and depicted in Fig.\ 2 such that
the medium is out-of-equilibrium for the period $0<t<t_0$, returns
to equilibrium at $t_0$ with an equilibration time constant
$\tau_{\rm eq}$, and stays in equilibrium for all times $t>t_0$:
\be V(t)=-T\Delta S =
\begin{cases} -V_0(1-e^{-t/\tau_{\rm eq}}) & (0\,,\,t_0),  \\  -V_0
e^{-t/\tau_{\rm eq}} & (t_0\,,\,\infty).
\end{cases} \ee $V_0$ is the information well depth as shown in Fig.\ 2.
So for times $t>t_0$, the quantized $|I_n|$'s are obtained by
solving \be \int_{t_n}^{t_0} [I_n + V_0(1-e^{-t'/\tau_{\rm
eq}})]dt' + \int_{t_0}^{t} [I_n + V_0 e^{-t'/\tau_{\rm eq}}]dt'  =
\tilde{n}, \ee where the time  $t_n$ (when the state first
appears), is given by $V(t_n)= -V_0(1-e^{-t_n/\tau_{\rm eq}}) =
I_n$, which solves to give \be t_n = -\tau_{\rm eq}
\ln(1+\frac{I_n}{V_0}). \ee With the above expression for $t_n$,
the information quantization Eq.\ (26) becomes
\begin{align} I_n & (t+\tau_{\rm eq}) +  (I_n + V_0) \tau_{\rm
eq}\ln(1 + \frac{I_n}{V_0}) =  \tilde{n} - V_0t_0 \nonumber\\
&\quad  - V_0\tau_{\rm eq}(1+2e^{-t_0/\tau_{\rm
eq}}-e^{-t/\tau_{\rm eq}}).
\end{align} For vanishingly small equilibration times
$\tau_{\rm eq}/t_0\rightarrow 0$, the thermodynamic potential for
information shown in Fig.\ 2 reduces to a rectangular information
potential well, and the algebra is greatly simplified without
changing the results in any significant way. In this limit, Eq.\
(28) reduces to $I_n t = \tilde{n} - V_0t_0$ and the quasistatic
eigen-informations supported by the medium are given by \be
|I_n|=\frac{V_0 t_0 - \tilde{n}}{t}\ \ \ \ (t>t_0).\ee The great
surprise in Eq.\ (29) is that it identifies evolving information
states supported by the environment in spite of the fact that the
medium supporting them is in a state of thermodynamic equilibrium
(heat-death) for all times $t>t_0$. This is the part of evolution
missed by the second law of thermodynamics but captured by the
second law of quantum complexity. The information states arise
because of the nonequilibrium history of the environment. Eq.\
(29) with $I_n = 0$ gives the total number of information states
supported by the medium $n_{\rm tot} = 2V_0t_0/\pi\hbar$.

On substituting Eq.\ (29) in Eq.\ (23), therefore, the
complexities of information states after heat-death are found to
increase monotonically according to \be C_n(t) = e^{S_{\rm max} -
|I_n|/T}=C_{\rm max} e^{-\frac{(V_0t_0-\tilde{n})/k_{\rm B}T}{t}
}.\ee So underneath thermodynamic equilibrium, the quantum states
can continue to evolve into progressively harder-to-describe
configurations. This completes the derivation of the second law of
quantum complexity from the quasistatic information quantization
rule in equilibrium circumstances $t>t_0$. It must be pointed out
that the functional form of $C_n(t)$ in Eq.\ (30) is indeed
universal because it arises from the second integral in Eq.\ (26),
which does not depend on the shape of the information well. As a
way of illustration the time evolution of the deepest information
state $|I_1(t)|$ characterized by the highest information content
(or uncomplexity) is shown in Fig.\ 2. For times $t>t_0$,
$|I_1(t)|$ decreases as a power law $t^{-1}$ as all other
information states do. With time $t$ measured in units of $(V_0t_0
- \tilde{n})/k_{\rm B}T$, the time evolution of complexity
$C_n(t)$ takes a common form applicable to all states \be C(t) =
C_{\rm max} e^{-1/t} .\ee Eq.\ (31) has a late-time
$C(t\rightarrow\infty) = C_{\rm max}$ behavior with an algebraic
$(1-t^{-1})$ approach to saturation, which is a much more detailed
prediction than the assumed linear growth \cite{CV}, although it
also predicts an intermediate linear regime with nearly constant
growth rate. To see this consider the rate or `speed' at which
complexity is increasing \be v(t)\equiv\frac{dC(t)}{dt} =C_{\rm
max}\frac{\exp(-1/t)}{t^2}. \ee $v(t)$ decays as $t^{-2}$ in the
long-time limit, which differs from the constant speed picture.
However, to see why previous results were roughly consistent with
a linear growth \cite{Linear,HJU}, we note that $v(t)$ has a
maximum when \be \frac{dv}{dt}=0 \ \ \Rightarrow \ \ t_{\rm int} =
1/2. \ee In a range of times around this intermediate time $t_{\rm
int}=1/2$, therefore, $v(t) \approx v(t_{\rm int})= 4/e^2$ is
roughly constant. So there is a broad region around which $v(t)$
changes slowly. It therefore appears that linear growth is an
intermediate-time approximation, and not a long-time  asymptotic
behavior that must indeed lead to saturation. This is a unique
prediction of the present theory that can be tested against model
simulations.

Finally, we note that for the out-of-equilibrium period $0<t<t_0$,
the quasistatic information quantization rule predicts gradually
increasing information eigenvalues in the time interval $\tau_{\rm
eq} \ll t < t_0$ given by \be |I_n(t)| = V_0-\frac{\tilde{n}}{t},
\ee meaning that in out-of-equilibrium circumstances it is rather
the uncomplexity that may grow with time, saturating to a value
determined by the deviation of environment from equilibrium $V_0$,
as also shown in Fig.\ 2 for $|I_1 (t)|$. Clearly, nonequilibrium
preparation can increase the uncomplexity and restore the
computational resource of a system \cite{nonequilibrium}. A
similar scenario was studied in details in Ref.\ \cite{DSZ} to
account for the rise of information and order in nonequilibrium
environments within the expanding universe.

In summary, the conjectured second law of quantum complexity was
derived from the quasistatic information quantization rule, thus
identifying a time evolution in equilibrium that although
different but includes the assumed linear growth as an
intermediate-time approximation, and not a long-time asymptote
that must also include a transition to saturation. In
nonequilibrium circumstances instead of complexity the
uncomplexity can grow, and thus replenish the computational
resource of a computationally spent quantum system. Furthermore,
an information-theoretic reinterpretation of the principle of
least action was developed consistent with Prigogine's principle
of minimum entropy production, which speaks of the natural
tendency of physical systems to spend their information resource
economically. This should serve to further elucidate the
fundamental role played by information in physical laws as
emphasized by the late Wheeler's famous conjecture `It from Bit'.





\end{document}